# Designing Internet of Behaviors Systems

Mahyar T. Moghaddam
*University of Southern Denmark*
Odense, Denmark
mtmo@mmmi.sdu.dk

Henry Muccini
*University of L'Aquila*
L'Aquila, Italy
henry.muccini@univaq.it

Julie Dugdale
*University Grenoble Alps*
Grenoble, France
julie.dugdale@imag.fr

Mikkel Baun Kjærgaard
*University of Southern Denmark*
Odense, Denmark
mbkj@mmmi.sdu.dk

*Abstract*—The Internet of Behaviors (IoB) puts human behavior at the core of engineering intelligent connected systems. IoB links the digital world to human behavior to establish human-driven design, development, and adaptation processes. This paper defines the novel concept by an IoB model based on a collective effort interacting with software engineers, human-computer interaction scientists, social scientists, and cognitive science communities. The model for IoB is created based on an exploratory study that synthesizes state-of-the-art analysis and experts interviews. The architecture of a real industry 4.0 manufacturing infrastructure helps to explain the IoB model and it's application. The conceptual model was used to successfully implement a socio-technical infrastructure for a crowd monitoring and queue management system for the Uffizi Galleries, Florence, Italy. The experiment, which started in the fall of 2016 and was operational in the fall of 2018, used a data-driven approach to feed the system with real-time sensory data. It also incorporated prediction models on visitors' mobility behavior. The system's main objective was to capture human behavior, model it, and build a mechanism that considers changes, adapts in real-time, and continuously learns from repetitive behaviors. In addition to the conceptual model and the real-life evaluation, this paper provides recommendations from experts and gives future directions for IoB to become a significant technological advancement in the coming few years.

*Index Terms*— IoB, Intelligent Systems, IoT, Human Behavior Modeling, Software Engineering, HCI, HITL, Cultural Heritage, Crowd Management.

## I. INTRODUCTION

The rise of intelligent connected socio-technical systems that perceive and act on human behaviors necessitates an approach beyond what is currently proposed by academia and industry. Human involvement in future intelligent systems' technological development will touch 40% of the world's population by 2023 [1]. The *Internet of Behaviors (IoB) links the digital world to humans*, their characteristics, goals, and interactions, and provides a desirable adjustment or trade-off between humans' quality of experience (QoE) and the system's quality of service (QoS). IoB can observe human behaviors, adapt itself accordingly, and continuously affect humans' decisions in implicit and explicit ways. The idea of IoB is formed around the principles of *i)* designing intelligent connected systems, such as Internet of Things (IoT) infrastructures, that monitor and predict human behaviors; *ii)* adapting the system to real behaviors; and *iii)* impacting humans' decisions and actions within a loop. IoB considers humans as a source of change and consequently computation. It does not only mean using mobile sensors (such as smartphones, watches, and other wearables) worn by people, but it also sees humans as the source of system intelligence [2].

IoB is hidden in many large IT companies' philosophy, where they aim to predict human behavior and its interplay with their systems. Amazon prepares for load peaks before the Christmas break with shopping behavior models; Netflix has sophisticated algorithms tuned based on human movie selection behavior that power their recommender systems; the Booking.com system tries to influence humans based on their trips history. Figure 1 shows a mixed human-robot production environment from the industry 4.0 ( I4.0) manufacturing setup at the University of Southern Denmark, which is monitored by an IoT system. The six-axis collaborative robot arms collaborate to perform assembly tasks for building a simplified drone. The mobile robots carry components from the operation table to automated storage. Similar to most intelligent systems, the *architects often do not consider the impacts of humans' dynamic behavior on the system and vice-versa*. IoB goes a step beyond what is considered as *human* by current software engineering (SE), human-computer interaction (HCI), and human-in-the-loop (HITL) communities. In those domains, humans are usually seen as *users* who use hardware and software components in order to interact with a system or *developers, engineers* and *clients* who collaborate and develop the system.

In the example shown in Figure 1, *SE* sees humans as being the developers who collaborate in programming the robots' behavior to deliver high-quality software in line with the requirements set by the lab's director. *HCI* considers the interaction between humans and computers (or robots in this case), and it tries to improve the system and users' performance with an optimized interaction design. *HITL* combines human intelligence with the I4.0 system's intelligence for better decision making. HITL directly involves users in the decision-making process, for example, scheduling the production line, where an interaction between the system and the user is still necessary.

IoB provides us with a more inclusive vision of humans that is different from other disciplines mentioned above. Humans are considered as citizens, occupants, or building visitors who *do not directly interact with the system* and are mostly *unaware that they are part of the system,* but who could *benefit* from the way the system works. After analyzing human behaviors, IoB could not only directly optimize the system's behavior, but also could silently impact humans' decision-making process, pushing them towards specific behaviors, moderating undesired behaviors, and improving their quality of experience. IoB provided us with awareness of human-related contextual knowledge, i.e., the situation and behavior of humans at a specific time and place. In order to act on such data, intelligent connected systems must be able to interpret possible human behavioral patterns and react to them, which requires comprehensive knowledge in the design phase.

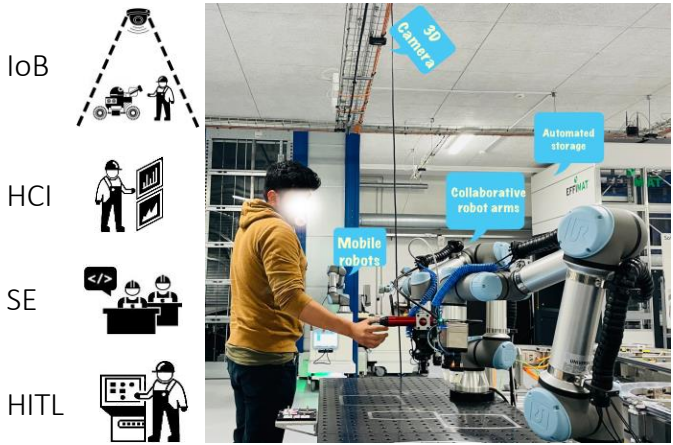

Fig. 1. IoB within a manufacturing domain.

By integrating *IoB* into our I4.0 lab, the production line behavior can be changed based on real and expected human behaviors. For instance, without IoB, the system can assure safety by inactivating the robot when humans come close. With IoB, the robots could predict people's future actions and calculate the safety risk to, e.g., change their movements instead of being inactivated. This makes an optimum trade-off between the QoS and QoE.

The questions and contributions that we address in this paper are:

**RQ1.** *What knowledge gaps exist in intelligent connected systems that necessitate the new concept of IoB?* For this, we reviewed the relevant literature to understand the needs and potential scientific value of IoB (*Section II*).

**RQ2.** *How can a comprehensive conceptual model, which includes quality concerns in various IoB related domains, be designed?* To answer this question, we conducted semi-structured interviews with domain experts from academia and industry. From this, we designed a well-grounded IoB conceptual framework focusing on the bidirectional impact of human activities and the system. The conceptual framework includes information about trends, design and privacy challenges, and future solutions (*Sections III and V*).

**RQ3.** *How can the IoB be used in practice in the software design and development phases?* We applied the IoB concept to design and adapt real systems software for crowd management of the Uffizi Galleries in Florence, Italy (*Section IV*).

This paper is organized as follows. The literature review is presented in *Section II*. *Section III* details the findings from expert interviews and presents the IoB model. *Section IV* describes applying the IoB approach to the Uffizi Galleries.

*Section V* provides some recommendations based on lessons learned. *Section VI* specifies threats to validity, and *Section VII* concludes the paper.

## II. LITERATURE REVIEW

IoB goes a step beyond how current software engineering, HCI, and HITL communities consider *humans*. As shown in Table I, in **SE**, humans are typically viewed as being developers and clients. Research interests in this area focus on the way they collaborate and interact in order to improve the quality of the software. Software development is hugely dependent on people [3] and is being studied by social, organizational, and psychology scientists. Wohlin et al. [4] discuss people in the software development process on the three levels of individual, unit, and organizational, which can determine the success of projects. The concept of behavioral software engineering (*BSE*) [5] addresses the cognitive, behavioral, and social aspects of software development towards a company's financial success.

TABLE I
IoB IN COMPARISON TO HCI, SE, AND HITL.

| Point of view | Specifications | | |
|---|---|---|---|
| | Human | Methods | Application Contexts |
| Human Computer Interaction | Users | User model, Task model, User evaluation | Interactive systems |
| Software Engineering | Programmers / Clients | Developers team building, Distributed cooperation, Agile development, Gender equality | Quality of software product |
| Human in the Loop | Operators / Supervisors | Performance assessment, Artificial intelligence, Machine learning | Interactive automation |
| IoB | Citizens / Occupants / Customers / Market | Sensors, Controllers, Actuators | Internet of Things |

**HCI** focuses on the interaction of *users* with computers and in developing interfaces to ease and enhance such interaction [6] [7]. HCI gives a high priority to users, their needs, and limitations [8]. Therefore, in user-centered design, the focus is on developing a user model (which characterizes a typical user), a task model (which describes the tasks that a user will undertake), and user evaluation (which assesses the efficiency, ease of use, usability and overall experience of the system's interface). While usability and user experience are addressed in software development [9], socio-technical theory [10] evaluates the usability concept in its social human and environmental context. Some researchers design their smart interactive systems based on HCI models to involve users in designing human-building interaction [7].

**HITL** is defined as a model that requires human interaction [11] and conforms to human factors impacting the system. The HITL approach combines human intelligence with machine intelligence, mainly for decision making. In intelligent systems such as IoT, HITL utilizes edge devices to sense and infer users' states [12] [13]. The system then uses such states to produce actions and adapt its behavior to match the users' state [14]. While HITL sees humans as an internal part of the system, it again requires human-system *interaction* [11]. The HITL approach re-frames an automation problem as an HCI design problem. While it may be advantageous for AI-based interactive systems, especially safety-critical systems, that are subject to oversight and benefit from human involvement, it does not emphasize the role of individuals and society. The assumption that the regulation and automation problem is solved by putting humans within the loop could be superficial in complex systems.

In our IoB concept, both system's and human behaviors are drivers of adaptation in order to improve both QoS and QoE. We are interested in the activities of citizens and occupants who are mostly unaware that they are part of the system. In IoB, the social environment is interlinked with a software system through intelligent elements. Although IoB systems are predominately associated with the intricacy of new technology, ignoring human behavior's complexity can cause problems. Humans may show fluctuating, unpredictable and irrational behaviors due to their individual, social, or environmental situations. Lack of consideration for those human-related dynamics can lead to system failure [15], low quality, or discomfort for people. Imagine, for example, an IoT-based crisis handling [16]–[18], crowd monitoring [19], [20], or queue management system [21] that pays little attention to humans' dynamic behavior [22]. This can result in system failure and inadequate quality and imposes safety risks and discomfort for people, who are the main reason for creating the system. Therefore, we argue that moving to IoB concept and techniques is essential.

### III. Conceptualization of IoB

The first draft of the IoB metamodel was designed based on our experience and the related literature. Then we conducted interviews to refine the model.

#### A. Semi-structured Interviews

A semi-structured interview [23] is a meeting in which the interviewers ask open-ended questions. This technique encourages free discussions in the direction the interviewee wishes to go, instead of using all pre-scripted questions. We chose this method since IoB is a new concept, and its characteristics should be discovered upon discussion with experts from various domains with different scientific languages. We followed four steps:

*1. Questionnaire Design and Process Specification.* We designed a questionnaire and detailed the interview process to specify the steps to be taken and points to bring up during the interviews. We informed the interviewees about the high-level topic of the interview one week in advance. We did not send them any guide or question to avoid any bias. All interviews were carried out in English. We targeted interviews of around 45 minutes. The interviews were conducted remotely via the Zoom platform. We asked interviewees' for their consent to record their videos for further analysis.

*2. Pilot Interviews.* Pilot interviews were then performed to assess the interview process and test the clarity of the questions and data extraction. The pilot interviewees were our colleagues who were not involved in the project. This step improved our approach, especially in managing and expanding fruitful discussions.

*3. Experts' Interviews.* Twenty top-rated scientists/practitioners (from 16 universities and 10 countries) were chosen from the following domains: software engineering, HCI, agent-based modeling, and cognitive science (see Table II). We chose these experts because of their specific interest and outstanding profile in various aspects of human involvement in IoT design and development.

*4. Data Analysis.* The detailed notes taken during the interviews by the interviewers were completed with the audio recordings that were later transcribed. The gathered data were then analyzed according to an open coding approach [24]. Based on the codes, we compared and tuned the results of the academic literature and the interviews. The outcome of this step helped us finalize the IoB conceptual model and specify its characteristics.

TABLE II
THE INTERVIEWEES (ORDERED ALPHABETICALLY).

| | Name | Country | University/Company | Role | Area of Expertise |
|---|---|---|---|---|---|
| 1 | Frederic Amblard | France | Toulouse 1 University Capitole | Professor | Agent-based Social Simulation |
| 2 | Jesper Andersson | Sweden | Linnaeus University | Professor | Software Engineering (Self-Adaptive Systems) |
| 3 | Elise Beck | France | University Grenoble Alps | Assoc. Professor | Multi-agent Systems - Geomatics |
| 4 | Jan Bosch | Sweden | Chalmers University of Technology | Professor | Software Engineering (Architecture, Testing) |
| 5 | Barbora Buhnova | Czech Republic | Masaryk University | Assoc. Professor | Software Engineering (Architecture) |
| 6 | Radu Calinescu | UK | University of York | Professor | Formal Methods for Adaptive Software Systems |
| 7 | Ivica Crnkovic | Sweden | Chalmers University of Technology | Professor | Software Engineering (Architecture) |
| 8 | Sophie Dupuy-Chessa | France | University Grenoble Alps | Professor | Human-Computer Interaction |
| 9 | Schahram Dustdar | Austria | Vienna University of Technology | Professor | Distributed Systems (Human-based Computing) |
| 10 | Benoit Gaudou | France | Toulouse 1 University Capitole | Assoc. Professor | Multi-agent Systems |
| 11 | Ilias Gerostathopoulos | Netherlands | VU University Amsterdam | Assist. Professor | Software Engineering |
| 12 | Rick Kazman | US | University of Hawaii | Professor | Software Engineering (Architecture) |
| 13 | Nenad Medvidović | US | University of Southern California | Professor | Software Engineering (Architecture) |
| 14 | Birgit Penzenstadler | Sweden | Chalmers University of Technology | Assoc. Professor | Software Engineering (Sustainability) |
| 15 | Jari Porras | Finland | Lappeenranta University of | Professor | Software Engineering (Sustainability) |
| 16 | Gustavo Rossi | Argentina | National University of La Plata | Professor | Context-awareness Systems |
| 17 | David Rozier | France | Cybersecura | Company | Cybersecurity & Data |
| 18 | Marjan Sirjani | Iceland | Reykjavik University | Professor | Software Engineering / Formal Methods |
| 19 | Marlene Villanova-Oliver | France | University Grenoble Alps | Assoc. Professor | Spatio-Temporal Semantics / Cognitive Science |
| 20 | Eoin Woods | UK | Endava | Chief Technical Officer | Software Engineering (Architecture) |

#### B. Designing the IoB Model from Interviews

The metamodel (Figure 2) depicts the vital concepts of systems, humans, and their interaction as a context for un-derstanding human involvement in the software design and adaptation process. Thus, the metamodel is divided into three parts: the left side of Figure 2 depicts the *human components*; the right side deals with the *system components*; the middle part shows the *interrelation* between system and humans.

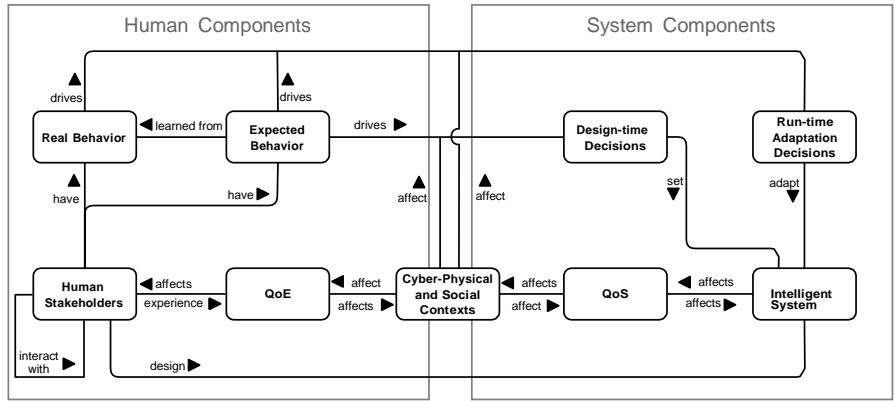

Fig. 2. IoB conceptual model.

**Human Components**

*Human Stakeholders.* The software system **stakeholders** may be users (if we follow the HCI and HITL interpretation), developers/clients/managers (software engineering interpretation), and citizens/occupants (IoB interpretation). In our I4.0 production line example, stakeholders are (Figure 1), developers, operators, managers, visitors, and final customers. The proposed IoB definition of humans is concerned with citizens'/occupants' individual *characteristics*, *behaviors*, and *goals and intentions*. Human characteristics include static and dynamic aspects such as age, gender, disability, social links, and profile; these can obviously vary from person to person. These characteristics influence humans' goals and intentions. For instance, in the Covid 19 context, we observe that young people (characteristic) are more likely to socialize (goal), while older people prefer to stay home and respect safety measures. Thus, humans' goals and intentions refer to what they want to achieve from the system.

Our interviewees stated that stakeholders could be classified into *active* or *passive*. Active stakeholders dynamically impact the system, while passive stakeholders are impacted by the software system. For instance, in an outdoor monitoring and rescuing system, the mission manager may use a dashboard to obtain situational awareness, a rescuer uses his mobile device to find the location and situation of an endangered person, and a victim indirectly uses the service provided by the software system. Furthermore, the experts emphasized that while humans are the focus, they are not the only stakeholders of the system. In the agent-based social simulation modeling community, some intelligent entities such as robots and autonomous cars may have human-like intelligent behaviors (*animate agents*) and could therefore be considered as stakeholders. There are other *inanimate agents* like buildings in an agent-based environment, but they are considered as part of the *context* since they have no intention and behavior. An improvement that came from the interviews and shown in Figure 2 is that *stakeholders can interact with each other*. This highlights that, while the interrelation among humans and systems is fundamental in IoB, the dynamics among humans themselves and the interaction of intelligent entities impact the behaviors. Thus, we emphasize that our model sees humans at both the individual and social levels.

Discussions with the experts led to simplifying the metamodel by merging stakeholders, human goal, intention, characteristics, and behavior into a single component called *human stakeholders*. This reduced the confusion caused for some communities. The original naming came from the agent-based social modeling and simulation community, who often use the Belief-Desire-Intention architecture [25] to model agent's cognitive reasoning of events and locations, their motivational state, and their deliberative state.

*Expected and Real Behaviors.* Human goals and intentions determine their behavior, e.g., only visiting the famous paintings in a museum or setting a specific robot's properties in a manufacturing environment. IoB is not only interested in behaviors but also in understanding what shapes and impacts the behavior (i.e., characteristics such as personality, past experience, and culture). This allows IoB to control those primary dimensions to influence behaviors. Many intelligent connected systems are solely based on **expected behaviors**, i.e., what the developers think or what the system predicts that people will do. As in cognitive engineering, we make a distinction between expected behavior and **real behavior** [26]. It is possible to determine expected behavior based on real behavior. In the example shown in figure 1, a data-driven or agent-based modeling approach could specify the movement trajectory of operators in the space. A robots routing algorithm can consider those trajectories and respect the safety measures whilst continuing with the tasks. However, if an operator moves unexpectedly, the robot behavior will be adapted based on observed real behavior. The above example is in line with the experts' view, highlighting that the real behavior is what is observed at run-time and could differ from the expected behavior. The important point is that we do not always need the same kind of adaptation based on observed behavior since, as a Cybersecurity expert pointed out, the behaviors could be malicious. The behaviors may concern legitimate direct/indirect use of the system, as well as disallowed use. Hackers use systems in a way that they are not supposed to be used; this behavior may still lead to architecture adaptation but in a positive way to enhance security.

Most interviewees mentioned different methods to predict, observe, model, or analyze human behavior in the design phase. Using questionnaires, humans describe some behaviors, but their descriptions may be prone to biases. Direct observation of behaviors can also be problematic since the chosen sample may be biased, and certain groups may have been ignored. Furthermore, some observed behaviors may not be reliable because irrational behaviors may not be normally shown, or some emotions are not aroused. Agent-based simulations are another approach to model potential human behaviors. However, the results can sometimes be unreliable due to imprecise parameter settings, unrealistic assumptions, and the difficulty of modeling complex behaviors.

Some software architects obtain information regarding expected behaviors from user experience designers who assess user behavior through usability, usefulness, and desirability when users interact with a product. Other software architects estimate expected behaviors based on previous experience of designing similar systems or using the datasets available from previous human-machine interactions. Data-driven approaches generally work well if the system to be designed is not completely new. However, if the expected behavior is significantly different from real behavior, they should either redesign or adapt the system. Designing self-adaptive systems are the best choice but requires a huge effort and precision in the design phase.

People working on computer simulations of human behavior stated that they collect data from interviews, workshops, government databases, social networks, and sensory data gathered by, e.g. a security camera. These data are then used to build more realistic models and simulators of human behavior. Some software engineers use Alpha/Beta testing for a randomized experiment with two variants and measure the performance difference. Software architects use human-generated data and history-based techniques to know what humans did while interacting with the system. They also interact with, e.g., user experience designers to receive user-related information to map that within the architecture. In HCI, the researchers use *implicit* or *explicit* methods, i.e., gathering data directly by asking users, or collecting data from interaction such as face detection, heatmaps, click precision, eye-tracking, text, and voice, etc. The interviewees' comments led us to assess the model from a more practical point of view as to how data on human behavior was collected and used by related disciplines.

*Quality of Experience.* QoE concerns assuring that users' have positive feelings when they directly or indirectly use a system. That could be developing techniques to quantify or qualify people's preferences and choices in different contextual situations. If our designs do not highlight QoE, our systems will not be responsive to their users. A provider needs to be able to observe and react quickly to quality problems, ideally before the customer perceives them [27]. Such quality concerns emerged the concept of QoE, which combines user perception, experience, and expectations with non-technical and technical parameters.

While context is the primary driver, human characteristics also play a crucial role. In a museum, visitors may be divided into different categories, e.g., those who visit selected artworks, walk close to the painting, and avoid congestion. For each of these categories, the QoE would be perceived differently. Experts suggested that we should consider individual and collective behaviors such as speed and vision variations, grouping, social attachment, Covid 19 social distancing, and emergencies that may significantly impact the QoE.

Based on experts' opinions, **privacy** is a significant challenge in QoE, and it is a priority to protect people's data while still providing behavior-oriented services. IoB is more interested in behaviors than individuals themselves. The interviewees noted that IoB should be equipped with concepts, methodologies, and tools to guarantee anonymity. At the architectural level, data processing on edge and fog without sending the data to the cloud helps to preserve privacy. There are also formal solutions to use the authorized data (by law). However, in IoB, the interaction between people and the system is minimal; therefore, it is better to separate behavior data from personal identities. An important aspect is increasing the acceptability of IoB within societies so that people perceive its benefits and feel that it is worth giving some personal data to receive a personalized service.

Other solutions that our interviewed privacy experts suggested are based on General Data Protection Regulations (EU GDPR) [28]: *i)* asking consent from all humans who are directly and indirectly using IoB. This solution is not desirable since the interaction between the system and the individual should be kept to a minimum with many intelligent systems; *ii)* aggregating data so that an individual's data cannot be identified. However, this depends on the context and situation, e.g., identifying individual older people in a huge crowd in a museum is difficult, but it would be easy in a restaurant with only three customers; *iii)* showing that the data are used for legitimate reasons that are critical for human safety, e.g., using a person's body temperature for Covid 19.

**System Components**

*Intelligent connected systems.* Intelligent system has functional and non-functional goals and requirements [29], which specify the system's functionality and quality under various contextual constraints and guarantee the system's operation. The environment and its contextual variability might also influence the **architecture** of the system. The architecture design [30] is also affected by the desired functional and non-functional requirements. The architecture includes a set of *components* that are bounded by *connectors* based on specific rules and constraints.

Experts emphasized that constraints imposed on the system by the operational environment to the architecture should be considered. Such constraints (that are a part of the context component) may be due to the availability of IoB resources (sensors, network, processors, actuators). Constraints may also include time aspects in real-time critical systems. Regarding the architecture, experts stated that the IoB architecture should include additional modules: *i)* a modeling module that helps to code different patterns of behaviors to prepare the running system with already generated plans, *ii)* a behavior decoder that keeps the identity and behavior separate to discover the required services based only on behaviors, *iii)* a service matching component that associates services with the captured behaviors [2].

Since human behavior is highly dynamic and complex, it makes architectural adaptation necessary and frequent. Humans may describe or exhibit some behaviors (while interacting with the user experience designers), which is very different from what they really do. This may result in the designed architecture not satisfying the intended goals or requiring major adaptation effort. Some experts also mentioned the difficulties they face when designing the software architecture in a collaborative way, interacting with other architects. Industrial experts stated that since the architecture level adaptation [31] occurs infrequently (because industry adaptation is more at the application level), considering imprecise human behavior imposes a costly process of architecture redesign.

*Quality of Service.* QoS considers quality factors that could be measured, and used to define, design, and adapt systems. Since the QoS are specified by stakeholders (and dynamically evolved), it may involve non-uniform views. For instance, the telecommunication community focuses on service quality toward the end-user, the IoT community highlights the capabilities of the network to provide fast packet transfer, and the HCI community has a user-centric view [32]. In IoB, the main question is how QoS measurements and control could be related to human perception of a service. To choose proper measures to keep human-perceived service quality above an acceptance threshold, the system should translate QoS parameters into user-level QoE perception and vice versa through the context. The main quality requirements IoB should assure are not only performance [33], [34], energy efficiency [22], resiliency [35], interoperability, and dependability, but also usability, ease of use, and efficiency. In the usability engineering domain [36], reaction time thresholds for user perception is 100 ms. This is the boundary at which a user feels the system reacts instantaneously; with less than 1 s the user's train of thought is maintained, although a delay is perceived; less than 10 s keeps the user's attention while exceeding 10 s implies the risk of the user abandoning the activity. Enhancing this QoE is tied with the performance enhancement using, e.g., reconfiguration of processing elements.

*Design-time decisions and run-time adaptation decisions.* As shortly mentioned above, expected human behaviors could influence the architectural design decision, and run-time adaptation decisions should get input from both expected and real behaviors. Modeling systems similar to IoB is challenging, especially as systems and contextual dynamics should be supported at run-time. According to the feedback we obtained from software architects, there are various steps to take: *i)* expected human behaviors and other contextual information should be anticipated using precise modeling approaches; *ii)* the system and physical environment constraints should be precisely understood; *iii)* different architectural models that are associated with behaviors and context should be designed; *iv)* the composed behaviors and architectures generate patterns that will provide input to the knowledge base of run-time systems; *v)* the run-time phase adopts a feedback control loop (such as MAPE-K [37]) for adaptation. Based on monitoring and analysis of contextual and system data, the control mechanism makes decisions based on the content of the knowledge base, executing further adaptations if needed. Despite the process explained above, some researchers believed that since IoB systems are massive, distributed, and continuously evolving, a differentiation between design-time and run-time may not be feasible. They suggest continuous architecting of the system based on dynamic context and system requirements.

**Human-System Links**

*Cyber-Physical and Social Contexts.* The interrelation among *human* and *system* aspects structures the conceptual idea of IoB. From the interviews, we understood that, in addition to human behaviors, context impacts both design-time and run-time decisions. The *context* component that partially falls within both the system and human blocks (but does not belong to them) includes *physical, temporal, social, computational, historical,* and *profile* contexts:

- *The physical context* relates to information regarding the *physical environment*, such as geographic location, temperature, humidity, noise, and light. This contextual dimension refers to the real executing environment in which the human is located, and the system could be implemented in real life. For example, the physical context of manufacturing with robots differs from a museum; thus, it impacts the system's QoS and humans QoE differently.
- *The temporal context* concerns time-related information that can affect IoB. This has implications on both the human and system sides. The situation and intention of humans evolve over time based on the context. For instance, visitors may change their behaviors by rushing around to see popular artworks if they know that a museum is closing. Another example might be that the intention and behavior of a company's operator in the morning (when he starts the day freshly) are different from his behaviors in the late afternoon (when he is tired).
- *The social context* concerns the direct or indirect interaction of people with each other, with the system, or with objects located in the physical or virtual environment. IoB considers social context as a key driver of system design and devel-opment. People may be socially attached to their friends in a museum. Such attachment could impact not only humans' behavior and experience but also the systems' configuration and quality.
- *The computational context* is related to the external resources available for the system, such as computing resources, communication bandwidth, and storage resources. Computation resources have a direct connection with the system side of IoB.
- *The historical context* deals with historical data that can affect the interpretation of information or the system's operation. In IoB, both the humans and systems histories could be stored in a knowledge base to be used for, e.g., learning purposes. The prediction of operators' movement trajectories in a mixed human-robot environment could be an outcome of analyzing their mobility behavior history.
- *The profile* concerns an entity's preferences for the specific contextual dimensions. Users' profiles might help IoB in suggesting specific less-crowded venues to visitors to impact their decision-making.

A deep understanding of the context is essential for designing the right IoB infrastructure. The lack of a uniform approach for capturing information associated with the context makes it difficult to fully understand the context model's needs and design approach based on its main characteristics. As mentioned above, the context includes humans' environment; anything happening around humans impacts their intention and behavior. This could be the weather, a user interface, or other humans around. It is worth mentioning that the level of impact that context makes on human behavior is different case by case; for instance, a person's behavior may be very different if they are told that they have 1 minute to evacuate a building [38] rather than 15 minutes (temporal context).

***Other links.*** We emphasize that the *expected human behavior* can establish the *design-time* setting to determine the system architecture and discover potential adaptation needs. The expected behavior could be derived through field studies and simulations or learned from real behavior. The *run-time setting*, which is adjusted by *real behavior*, is the source of setting online self-adaptation. Several yet unsolved issues should be investigated, ranging from dynamic human behavior to the self-adaptation process.

Based on interviewees' comments, we added an impact feedback loop between the system and humans to reflect the idea that humans can impact the software design and adaptation and that the system can impact human behavior within the environment. For instance, in adaptive user interfaces [39], the system could capture humans' emotions and adapt the interface based on their feelings. This adaptation then again impacts the humans' emotional behavior. On the system side, the adaptation makes the dynamics of the system clear. As shown in Figure 2, we highlight human dynamic as well by *i)* specifying interaction among humans, *ii)* showing a bidirectional impact among QoE and humans and their behavior, *iii)* including the social environment and considering temporal and social contexts. As designers of IoB, we often consider that the system changes. However, we should avoid unrealistic assumptions that humans are always the same but that they can change behavior more frequently than the system changes.

## IV. REAL-LIFE APPLICATION

We applied our concrete IoB model to the queue management system of the Uffizi Galleries in Florence, Italy. Uffizi is one of the most visited museums in Italy, with over 2 million visitors every year. We updated the traditional ticketing system to one that tracks behaviors, models them, and together with environmental factors, makes data-driven predictions, and adapts to the system's changing conditions. Dealing with the museums' congestion is a socio-physical system where the dynamic interaction between visitors and decision-makers can be modeled. The duration of visits depends on the time of the day, the day of the week, and the season. Such aspects are highly impacted by *human characteristic*, such as being an individual or in a group. Groups of people typically move together because of solid social attachments, which may slow down the overall group walking velocity [40]. As shown in Figure 3, we designed an IoB-based software/hardware infrastructure to manage queues and continuously monitor humans' reactions to the system to be able to improve its functionality and quality. Dynamic data acquisition allowed us to design a model that can predict human mobility behavior. We applied each element of the IoB model to the Uffizi case.

**Humans in Uffizi.** In the Uffizi, all *stakeholders* have an interest in the improved queue management system. By implementing efficient queue management, managers maximize profit, museum professionals attain job satisfaction, and citizens enjoy a higher QoE. The most fundamental social and individual *behavior* is the action of people choosing to visit the gallery, particularly during the high season and on certain days of the week. 4/left lower shows the time series of daily visitors during our experiments. Humans may also have the specific *intention* to only visit particular artworks, which causes internal congestion in the museum. The *context* impacts the visitors *intention*, as seasonality causes an increase in visitor numbers in spring and summer. The effect of other *contextual* situations such as strikes, weather, and interruptions in the air-conditioning system are also significant. Some visitors may also have a firm *intention* to visit the museum on the first Sunday of the month when it is free. These variations were taken into account in our system. However, we cleaned and rendered the data to provide a more homogeneous statistical point of view.

The IoB metamodel shows how the context impacts human QoE and their goals and intentions. In the Uffizi, the duration of visits *(intention)* depends on the time of the day *(context)*, see Figure 4/lower middle. The average duration for early morning is more than 155 minutes, around 135 minutes for late morning, and 83 minutes after 16:00 hrs. This variability highlights visitors' mobility *behavior* during a day or the year. An important *human characteristic* that we considered is whether a person is in a group (a crowd of more than five persons) or not. Groups enter the museum and move together. We observed that, on average, 16 groups booked their tickets during a one-hour time-span. This is approximately 5% of the total bookings. Groups of people typically move together because of strong social attachments, which may slow down the overall group walking velocity [41].

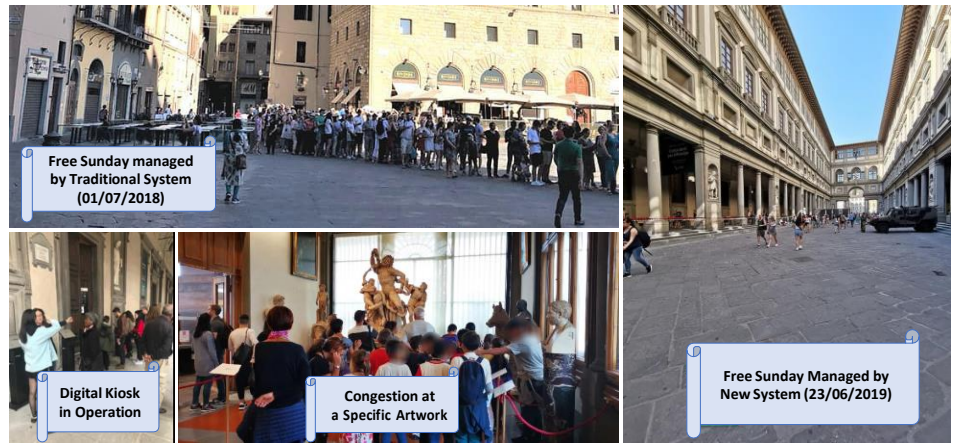

Fig. 3. The enhanced queue management system.

**System's Design for Uffizi.** We dynamically capture human mobility *behavior* in the Uffizi to provide a set of services that facilitate ticket booking and eliminate long queues. We designed an architecture (Figure 5) consisting of various sensing, computation, and actuation elements, which is adapted based on *design-time and real-time decisions*. Human *characteristics* (such as social group links); their *real behaviors* (such as live arrival rate at the museum, external waiting time, visiting time, and internal flow); and *contexts* (such as light) were captured using IoT sensors.

As shown in Figure 5 in addition to the RFID system, we used some of the 500+ already installed CCTV cameras and used people counters to get more precise data. RFID technologies require equipping pedestrians with unique tags. However, visitors wish to minimize their *interaction* with the software system and do not want to be constrained with any additional device. Hence, we obtained data by tracking people's mobility by mostly using CCTVs and counters. Data on human mobility is sent to a processing and storage component, a cloud, where it is analyzed. We used the MVC architectural design pattern, which was implemented using the Pivotal Spring framework. The Spring web application comprises an MVC controller, data service, and repository. The controller principally concerns an interface for data acquisition. The collected data is stored in a MySql DBMS. The solver is developed in Python by calling the ILP solver. The DBMS provided the solver inner-component with summary and real-time statistics.

The IoT system has an algorithmic core. The algorithm solver is a part of *context* and gets input from the *physical environment*. More specifically, the solver considers static data regarding the museum's capacities, real-time human *behavior* of a group or individual booking, booking time preference *(intention and characteristics in IoB)*, no-show *(behavior in IoB)*, and visiting time *(behavior in IoB)*. Such data passes through the Spring web application to the solver to calculate the available time slots for visitors every 15 minutes.

The IoT actuators are the kiosks and dashboards, which provide situational awareness for visitors and operators. The available time slots are shown to visitors in the kiosks where they can reserve their tickets. The kiosks receive the periodical input from the solver to print available tickets for time-slots. The dashboards let museum *managers and operators* monitor data acquisition, people counts, and visiting time, via an interactive dashboard. The dashboard also provides reliable information to the museum personnel on the number of *visitors* that can access the museum in a given time slot with little/no waiting time on a given day. This reduces the operators' effort and gives situational awareness on the *system* and *human* mobility. Our preliminary IoB HW/SW system provides easy and flexible booking to enhance the visitors' QoE. In the following subsection, human reactions to the new system and how the system was consequently adapted are discussed.

**Human-System Interrelation in Uffizi.** We collected data on people's reactions to the system. Our queue management solution has been deployed through the installation of outside digital kiosks (Figure 3). To decide on the number of required kiosks, we considered several parameters on *human behavioral reaction*: *i)* the time for individuals to interact with the interface and book their ticket (around 15 seconds), and *ii)* the maximum number of people arriving at the museum in each 15-minute time slot (which is 268). From this, we calculated that seven kiosks would keep the waiting time to less than 10 minutes in the worst case. The kiosks' functionality significantly enhanced the visitors' QoE. Once a visitor has booked their ticket, they can leave the museum area and enjoy the surroundings until her/his reservation time approaches.

Another consideration of how people behave was the number of *no-shows*; people book a ticket but do not show up to the museum. This may happen when people have free tickets and do not feel a commitment to attend. We monitored this no-show behavior and found that such behavior varies during the day as shown in Figure 4/upper. From the data, we observed that visitors who reserved their tickets in the morning for the evening tend to have a high no-show rate. In this case, we can decide to overbook some evening slots based on a no-show model. Such overbooking is handled by virtually increasing the museum's capacity and making more tickets available for slots where the no-show rate is expected to be high. The overbooking level was chosen by considering the worst case where the average time that people spend in the museum is reasonably high, and the no-show rate is relatively low.

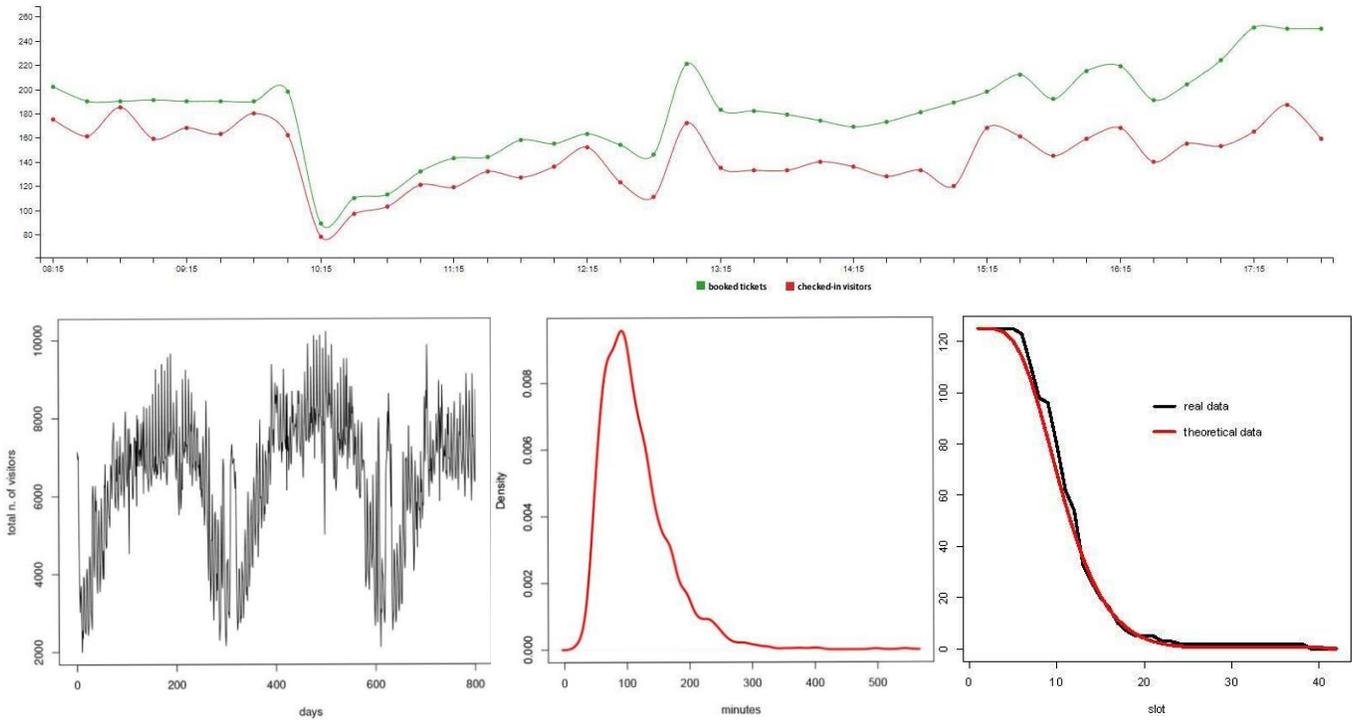

Fig. 4. *These are the last data we gathered before the Covid 19 pandemic.* **Top:** Ticket distribution and number of no-shows on January 6th, 2019. – **Lower left:** Time series of daily visitors during the whole experimentation period – **Lower middle:** the duration of visits on June 3rd, 2019 – **Lower right:** Comparison of live data and model predictions for the number of visitors leaving the premises as a function of exit time on June 2nd, 2019.

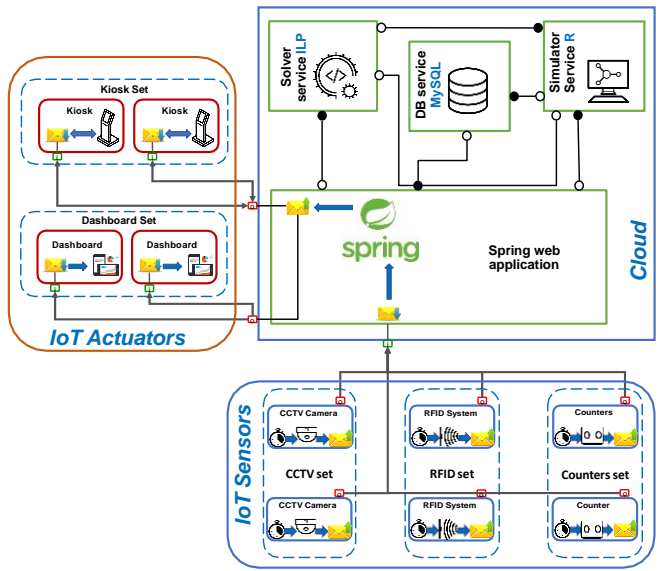

Fig. 5. Software architecture.

We further observed that the booking time in kiosks and the distribution of available tickets affect the no-show rate. Suppose visitors spend a relatively long time in the kiosk. In that case, their decision might be more considered due to a lower level of stress and consultation with other social group members. On the other hand, if after a statistical analysis of the no-show rate, the available tickets are distributed over a day (e.g., morning, afternoon), the time gap between a ticket booking and the visit will be reduced towards a lower now-show rate. Making the now-show rate as small as possible has a positive impact on QoE since more people will have the opportunity to visit the museum. Using our system adaptation, we observed a reduction in our no-show rate for five consecutive free visiting days, as shown in Table III.

Having the system adapt to human behaviors could lead to a learned *prediction* model. After performing a detailed data analysis on mobility behavior, we identified a fundamental parameter for queue management: *the duration of the visit to the museum*. For each time slot, we constructed an output probability vector based on the characteristic of that slot.

TABLE III
NO-SHOW COUNT FOR 5 CONSECUTIVE FREE VISITING DAYS, USING THE
SYSTEM ADAPTATION TO HUMAN REACTIONS.

| Date | Issued Tickets | No-Show | % No-Show |
|---|---|---|---|
| 2019-03-05 | 7496 | 1480 | 19.7 |
| 2019-03-06 | 7290 | 1360 | 18.9 |
| 2019-03-07 | 7214 | 1257 | 17.4 |
| 2019-03-09 | 7434 | 961 | 12.9 |
| 2019-03-10 | 7334 | 870 | 11.9 |

All of these vectors form a matrix where it is possible to predict the exit time, given a visitor's entry time. In this way, we dynamically model the number of people present inside the museum and determine the maximum allowed number of incoming visitors for each entry slot. Live data and our model predictions are compared in Figure 4/lower right. An important feature, which lends general validity to our model, is that the maximum allowed number of incoming visitors is a dependent function of the permitted maximum. Since this is true of many controlled access buildings, our approach could be widely applied.

## V. LESSONS LEARNED

Putting humans as the primary consideration is the most crucial aspect of IoB. We believe that applying IoB could bring economic, individual, social, and technical benefits to IoT systems. From the interviews and real application, we note that:

- Interviewees involved in the semi-structured interviews clearly recognized the role and benefit of taking into account human behaviors when designing intelligent systems' architecture.
- The human and system parts of the IoB model share common components (instantiated in the respective field), such as environment, context, behavior.
- Both human and system behaviors dynamically change based on the contexts.
- Social links and preferences should be seriously considered and modeled within IoB since they can impact the functionality and quality of the system.
- Capturing humans' behavior and adapting the system improves people's QoE and QoS.
- Providing situation awareness for stakeholders positively impacts their decision-making behavior.
- Adapting the interface characteristics (e.g., allowing a longer booking time in kiosks) based on human context and behavior (e.g., slow reaction time or need for collective decision-making) can significantly improve the performance of the system by decreasing the human errors and undesirable reactions (e.g., no-show).
- Collecting insights from users by, e.g., self-reporting methods gives a more robust and complete picture of the system. In the Uffizi case, we organized monthly meetings with the museum manager and operators to receive their feedback on the system performance. We are also continuously in touch with visitors and document their thoughts about the system to consider them a baseline for system improvement.

## VI. THREATS TO VALIDITY

*External validity:* In our study, the most severe threat to validity may be that the selected studies or interviewees do not represent a complete view of IoB. We mitigated this threat by *i)* applying precise criteria on the studies; *ii)* selecting the experts from both IoT and social/cognitive science.

*Internal validity:* This refers to the level of influence that extraneous variables may have on the design of the study. One internal validity is that most interviewees were selected from the academic community, which may neglect the experience of developers designing intelligent systems. We tried to mitigate internal validity threats by *i)* rigorously defining and validating the structure of our study, *ii)* orienting the questions to the practical side of IoB, *iii)* defining our classification framework by carefully following the coding process, and *iv)* conducting a well-structured vertical analysis of the results.

*Construct validity:* This concerns the validity of extracted data with respect to the research questions. We mitigated this in different ways: *i)* planning the study; *ii)* holding collaborative sessions for data collection and analysis; *iii)* unifying the interview using a guide. In interviews, we performed pilot interviews to minimize the biases.

*Conclusion validity:* This concerns the relationship between the extracted data and the obtained results. We mitigated this by applying systematic methods for coding, documenting, and analyzing the results.

## VII. CONCLUSION

We presented a conceptual framework for IoB systems design. The conceptual model was designed based on an exploratory study by synthesizing academic evaluations with experts' interviews. We evaluated the resulting approach by applying it to a real case. The approach strongly links the IoT system with human aspects and facilitates system design and development based on behaviors. The IoB model emphasizes improving the humans' QoE within IoT systems when human interaction is unnecessary but when people benefit from how a system works. It also deals with QoS issues linked with human behavior. We showed that IoB could improve IoT systems' functionality by continuous adaptation based on perceived human behavior. Humans also adapt their behavior and their context changes. This bilateral interaction and adaptation support a system that can predict real human behavior at real-time.

ACKNOWLEDGEMENT

This work is supported by the Innovation Fund Denmark for the project DIREC (9142-00001B).